\documentclass[11pt,twoside,a4paper]{article}
\usepackage{epsfig}
\usepackage{graphicx}
\usepackage{srt_style}
\def\solar{\ifmmode_{\mathord\odot}\else$_{\mathord\odot}$\fi~}
\def\hub{$H_0=100$ km s$^{-1}$ Mpc$^{-1}$, $q_0 = 0.5$}
\def\sprop{$S_{\nu} \propto \nu^{\alpha}$}

\def\deg{\ifmmode $\setbox0=\hbox{$^{\circ}$}$^{\,\circ}
          \else    \setbox0=\hbox{$^{\circ}$}$^{\,\circ}$\fi\,}

\begin{document}
\baselineskip .5cm
%
\title{The Role of Millimeter VLBI Observations in AGN Research}

\author{T.P. Krichbaum, A. Witzel, and J.A. Zensus}

\affil{Max-Planck-Institut f\"ur Radioastronomie, Auf dem H\"ugel 69, D-53121 Bonn, Germany}

%
\begin{abstract}
VLBI at millimeter wavelengths (mm-VLBI) allows the detailed
imaging of compact galactic and extragalactic radio sources with 
micro-arcsecond scale resolution, unaccessible by other observing techniques.
Here we discuss the scientific potential of mm-VLBI for present and future
research on `Active Galactic Nuclei' (AGN) and their powerful relativistic jets. 
With the new generation of large radio telescopes and interferometer arrays operating 
in the millimeter radio bands (e.g. ALMA), the ultimate vicinity of super massive Black Holes,
and eventually even their event horizon, could be imaged. With its large collecting area,
and in combination with these future telescopes, the Sardinia Radio Telescope 
could form the World's `sharpest' astronomical imaging machine.
\end{abstract}

\section{Introduction}
Even after more than 40 years after Marten Schmidt's discovery
of the cosmological redshift of the hydrogen lines in 3C\,273, and of astrophysical 
research on Active Galactic Nuclei (AGN) in general, 
the enigma of the origin of their extreme luminosity (ranging from radio to
Gamma-ray bands) and the creation mechanism for the highly relativistic plasma jets 
(often extending over many hundred kpc) still is not solved. Although the majority
of the scientific community regards accretion onto supermassive black holes as the
most plausible explanation for the `quasar phenomenon', 
many details of the astrophysical processes taking place in the centers of these most luminous 
objects in the Universe still remain unexplained.
In particular is the question of how the relativistic jets are made, accelerated and confined
not satisfyingly answered. In order to test existing theories (e.g. the Blandford-Payne
magnetic slingshot mechanism), which try to explain the accretion process and 
the subsequent energy release and jet creation, the direct imaging of the innermost 
regions of Active Galactic Nuclei is of great importance. The technique of 
interferometry is the only observing method, which leads to such direct images.

In {\bf V}ery {\bf L}ong {\bf B}aseline {\bf I}nterferometry (VLBI) the angular resolution
can be improved, either by increasing the distance between the radio telescopes or
by observing at shorter wavelengths. The first approach leads to VLBI with orbiting radio antennas 
in space (eg. VSOP, Hirabayashi et al. 2000), which however at present gives only an angular resolution
of 0.2--0.3\,mas (1\,mas = $10^{-3}$ arcsec) at 5\,GHz. 
The second possibility leads to VLBI at millimeter wavelengths (mm-VLBI),
which furthermore facilitates the imaging of compact structures, which are 
self-absorbed (opaque), and therefore not directly observable, at the longer centimeter wavelengths.
Nowadays, mm-VLBI observations are regularly performed at 86\,GHz ($\lambda=3.5$\,mm), where
images with an angular resolutions of up to $\sim 50 \mu$as ($1 \mu$as = $10^{-6}$ arcsec) are obtained.
The two upper parts of Table 1 summarize the stations, which participate in such global 3\,mm-VLBI 
experiments.
VLBI observations at even shorter wavelengths are technically possible 
(see e.g. Krichbaum et al. 1997 for VLBI at 215\,GHz; Greve et al. 2002 for VLBI at 147\,GHz),
but have not yet passed the stage of test experiments on more than on single baselines.
With the coming new generation of large millimeter telescopes and arrays (see
Table 1, lower section), 
global VLBI observations at 2 and 1\,mm wavelength should become possible within the
next few years. Such observations would result in images of up to $\sim 20 \mu$as resolution,
which in the case of the quasar 3C\,273 would correspond to a spatial resolution
of $\sim 400$ Schwarzschild radii (for a $10^9$\,M\solar black hole). One can therefore
imagine that for less distant objects (e.g. for nearby radio-galaxies like M\,87, Cyg\,A and 3C\,84,
or for the compact radio source Sgr\,A* in the Galactic Center), future ground based 1\,mm-VLBI
or space based 3\,mm-VLBI should allow the direct imaging of the vicinity of 
supermassive black holes and eventually even the event 
horizon\footnote{For Sgr\,A* the present upper limit to the source size
is $\sim 20$ Schwarzschild radii (cf. Krichbaum et al. 1998).}. Thus direct observational
tests of general relativity and space-time curvature near large masses will become possible.

In addition to the basic motivation of trying to image the close vicinity of a black hole
in an AGN with mm-VLBI, other questions related to the observed source activity 
(flux density variability, jet kinematics, broad band spectrum) are of direct interest. In 
jets of compact radio sources, mm-VLBI can detect structural changes at a relatively
high observing frequency and in a very early evolutionary stage. This can be used
to study the relation between flux density outbursts (broad band, from radio to Gamma-rays) 
and ejection of new jet components. Such `outburst-ejection' correlations seem to be 
present in many blazars and are a direct signpost of the acceleration mechanism acting at the center. 
Through its very high angular resolution, mm-VLBI images of radio-jets
also allow to locate moving features in the jet with very high accuracy. 
Jets of nearby ($z \leq 0.2$) galaxies can also be resolved transversely. All this facilitates
a more precise determination of the structural variations and the jet kinematics. 
Images obtained from mm-VLBI furthermore 
allow to trace the jet curvature, frequently observed in many AGN, much 
closer to the nucleus. Such curvature,
which increases towards the center, is an indicator of complicated internal plasma physics 
(e.g. hydrodynamic or magneto-hydrodynamic instabilities, oblique shocks) and possibly of
twisted magnetic fields anchored in a rotating accretion disk.
\begin{table}
\begin{center}
\begin{tabular}{|l|c|c|l|c|c|} \hline
Station     &  D    &  {\bf $T_{sys}$}&  Gain     &{\bf $\eta_A$} & SEFD \\
            &$[\rm m]$& $[\rm K]$     &$[\rm K/Jy]$&              & $[\rm Jy]$  \\ \hline \hline
                     &       &                 &           &               &     \\
Effelsberg, Germany  & 100   &     150         &  ~0.14    &  ~0.08        & 1070  \\
Haystack, MA, USA    & 37    &     150         &  ~0.058   &  ~0.15        & 2590  \\
Pico Veleta, Spain   & 30    &     120         &  ~0.14    &  ~0.55        &  860  \\
VLBA (7), USA        & 25    &     150         &  ~0.032   &  ~0.18        & 4680  \\
Onsala, Sweden       & 20    &     250         &  ~0.045   &  ~0.40        & 5550  \\
Sest, Chile          & 15    &     200         &  ~0.038   &  ~0.60        & 5260  \\
Mets\"ahovi, Finland & 14    &     300         &  ~0.017   &  ~0.30        & 17650 \\ \hline \hline
                     &       &                 &           &               &     \\
Nobeyama, Japan      & 45    &     150         &  ~0.17    &  ~0.30        & 880 \\ 
Ovro, CA, USA        & 6x10.4&     200         &  ~0.084   &  ~0.50        & 2380 \\
Hat Creek, CA, USA   & 9x6.1 &     200         &  ~0.050   &  ~0.55        & 4000 \\  
Quabbin, MA, USA     & 14    &     300         &  ~0.026   &  ~0.47        & 11540 \\ 
KittPeak, AZ, USA    & 12    &     150         &  ~0.026   &  ~0.64        & 5770  \\ \hline \hline
                     &       &                 &           &               &     \\
ALMA, Chile          & 64x12 &     100         &  ~1.82    &  ~0.70        &  55 \\ 
GBT, WV, USA         & 110   &     100         &  ~1.00    &  ~0.35        & 100 \\         
CARMA, CA, USA       &6x10+9x6&    100         &  ~0.14    &  ~0.55        & 710 \\
P. de Bure, France   & 6x15  &     100         &  ~0.18    &  ~0.51        & 560 \\ 
SRT, Sardinia        & 64    &     100         &  ~0.35    &  ~0.30        & 290  \\
LMT, Mexico          & 50    &     100         &  ~0.43    &  ~0.60        & 230 \\
Yebes, Spain         & 40    &     100         &  ~0.13    &  ~0.30        & 770 \\ \hline
\end{tabular}
\end{center}
\caption{
VLBI telescopes operating in the 3\,mm band (86-100\,GHz). The columns list
the station name and location (col. 1), the antenna diameter (col. 2), a typical system
temperature (col. 3), the antenna gain and aperture efficiency (col. 4 and 5) and
the system equivalent flux density SEFD (col. 6).
Antennas which regularly participate in global 3mm-VLBI campaigns are listed
on top. The second part of the Table summarizes telescopes, which either participated
in the past in such experiments, or which participate occasionally. The lower part of the Table
lists some antennas (planned or under construction), which could participate 
in mm-VLBI in the future. For these antennas, the aperture efficiencies and the gains
were roughly estimated based on data available in the internet and the literature.
The detection sensitivity on a VLBI baseline between two antennas can be estimated
by $\sigma_{\rm [Jy]}= \eta^{-1} \cdot \sqrt{{\rm SEFD}_i \cdot {\rm SEFD}_j / (2 B t)}$, were
$B$ is the observing bandwidth in [Hz], $t$ is the integration time in [sec], $\eta$ 
is the correlation loss factor (0.88 for 2 bit sampling) and the SEFDs are 
taken from column 6 for two antennas $i$ and $j$.}
\end{table}

\section{The inner jet of 3C\,273, a well studied case}
In the following paragraph, we present some results for the inner jet of the quasar 3C\,273, 
which can serve in this context as a typical example to illustrate the capabilities of mm-VLBI for future
AGN research not only in this but also in other objects:

At sub-milliarcsecond resolution, 3C\,273 shows a one sided core-jet structure of several
milliarcseconds length. The jet breaks up into multiple VLBI components, which -- when
represented by Gaussian components -- seem to separate at apparent superluminal
speeds from the stationary assumed VLBI core. The cross-identification of the model-fit components,
seen at different times and epochs, is facilitated by small ($< 0.2$\,mas), and to first
order negligible, opacity shifts of the component positions relative to the VLBI-core. 
Quasi-simultaneous data sets (cf. Fig.\ 1) demonstrate convincingly the reliability of the component 
identification, which results in a kinematic scenario, in which all detected jet components
(C6 -- C18) move steadily (without `jumps' in position) away from the core (Fig.\ 4).
For the components
with enough data points at small ($<2$\,mas) and large ($> 2$\,mas) core separations, quadratic fits
to the radial motion r(t) (but also for x(t) --right ascension, and y(t) --declination) represent
the observations much better than linear fits.
Thus the components seem to accelerate as they move out. The velocities range typically from
$\beta_{app} =3 -8$ (for \hub).

\begin{figure}[p]
\begin{minipage}[t]{6.0cm}{
\psfig{figure=kri_fig1.epsi,width=5.7cm}
{\footnotesize Fig.\ 1:
3C\,273 at 22\,GHz (top), 43\,GHz (center),
and 86\,GHz (bottom) observed in  1995.15 -- 1995.18.
Contour levels are -0.5, 0.5, 1, 2, 5, 10, 15, 30, 50, 70, and
90\,\% of the peak of 3.0 (top), 5.4 (center), and 4.7\,Jy/beam (bottom).
For the 22\,GHz map, the 0.5\,\% contour is omitted. The restoring beam is
$0.4 ~{\rm x}~ 0.15$\,mas in size, oriented at $\rm{pa}=0 \deg$. The maps are centered on the
eastern component (the core), the dashed lines guide the eye and help to identify
corresponding jet components. 
}
}
\end{minipage}
~~
\begin{minipage}[t]{6.5cm}{
\vspace{-16.5cm}
\psfig{figure=kri_fig2.epsi,width=5.4cm,angle=-90} 
{\footnotesize ~~\\ Fig.\ 2:
Spectral index variations along the jet. For 1997 (circles, solid lines),
the spectral index gradient is calculated directly from the intensity profiles of the
maps at 15 and 86\,GHz. For 1995 (squares, dashed line), the spectral indices were
derived from Gaussian component model fits at 22 and 86\,GHz. We note that 
during 1995 -- 1997, different jet components occupied this jet region.
The spectral profile along the jet, however, did not change very much. 
}
\vspace{1.0cm}

\psfig{figure=kri_fig3.epsi,width=4.9cm,angle=-90}
{\footnotesize ~~\\ Fig.\ 3:
The motion of the mean jet axis in the inner jet of 3C\,273 at 15\,GHz. Symbols denote
for different epochs: 1995.54 (circles), 1996.94 (squares), 1997.04 (diamonds),
and 1997.19 (triangles). The transverse oscillations of the ridge line are measured 
relative to a straight
line oriented along $\rm{pa}=240\deg$. We note the systematic longitudinal
displacement of maxima and minima. 
This corresponds to an apparent pattern velocity of $\beta_{app} \simeq 4.2$,
which is by a factor of up to 2 slower than the component motion.
}
}
\end{minipage}
\end{figure}

Dual-frequency maps obtained in 1995 (22/86\,GHz) and 1997 (15/86\,GHz) allow to measure
the spectral index gradient along the jet (see Fig.\ 2). The spectrum of the jet oscillates
between $-1.0 \leq \alpha \leq +0.5$ (\sprop).  Most noteworthy, the spectral gradients did not change
by much over the 2 year time interval, although different jet components occupied 
this region during this period. 
Therefore, the geometrical (eg. relativistic aberration) and/or the physical environment
(pressure, density, B-field) in the jet must determine the observed properties of the VLBI components.
Hence, we conclude that the latter do not form `physical entities', but seem to react to the physical 
conditions in the jet fluid and from its boundaries.

Further evidence for a fluid dynamical interpretation comes from a study
of the shape of the mean jet axis and the transverse width of the jet. Both oscillate
quasi-sinusoidally on mas-scales. At 15\,GHz, the variation of the ridge-line with time 
could be determined from 4 VLBA maps obtained during 1995 -- 1997 (Fig.\ 3). 
The maxima and minima of the ridge-line are systematically displaced. This `longitudinal'
shift suggests motion with a pattern velocity of $\beta_{app} =4.2$, about a factor of 2
slower than the maximum observed jet speed ($\beta_{app} \leq 8$). 
The sinusoidal curvature of the jet axis, however, could also indicate jet 
rotation rather than longitudinal waves. 
Helical Kelvin-Helmholtz instabilities propagating in the jet sheath (cf. Lobanov \& Zensus 2001) 
could mimic such rotation, which, when seen in projection, would also explain 
the spectral index oscillation seen in Figure 2.
\begin{figure}[t]
\centerline{\psfig{figure=kri_fig4.epsi,width=9cm,angle=-90}}
{\footnotesize Fig.\ 4:
Relative core separations r(t) for the components C6 -- C18 plotted versus time.
The legend on the right identifies symbols with VLBI components. The lines
are least square fits to the data. Note the acceleration of the motion. For the
inner and newest jet components ($r < 1$\,mas, $t > 1996$) the data do not yet allow
to discriminate between constant and accelerated motion.
}
\vspace{-0.7cm}
\end{figure}

Additional support for a rotation of the jet around its axis 
comes from the study of the ejection of new
jet components, appearing shortly after prominent flux density outbursts.
From Figure 4 and from the VLBI maps, we determined the velocity and the direction
of ejection of such new jet components, by fitting a straight line 
to the inner region of the jet (for $r \leq 0.5$\,mas).
From these fits we also obtained the time of ejection, which
correlates remarkably well with the occurrence of the known millimeter/optical
outbursts. In Figure 5 we plot the ejection velocity and its direction (position angle) versus the time
of ejection ($t_0$). It is remarkable that ejection velocity and ejection direction
show systematic, quasi-periodic
variations, indicating a precession of the footpoint of the jet with a period of
15--16\,yrs (see also: Abraham \& Romero, 1999). 
In 1982 -- 1997 a correlation between optical flux and ejection velocity
is indicated (Fig.\ 5, right panel). At earlier times such a correlation is less obvious,
however can not be excluded, since at these times the accurate determination 
of the ejection velocity was limited by the quality and time sampling 
of the early (pre 1990) VLBI data. Qian et al. (2001) explain the correlation of
ejection velocity and optical flux by a periodic modulation of the jet flow Lorentz 
factor in a binary black hole system with a precessing accretion disk.

\begin{figure}[t]
\begin{minipage}[t]{6.0cm}{
\psfig{figure=kri_fig5a.epsi,width=5.0cm,angle=-90}
}
\end{minipage}
~~
\begin{minipage}[t]{6.5cm}{
\psfig{figure=kri_fig5b.epsi,width=5.0cm,angle=-90}
}
\end{minipage}
{\footnotesize ~~\\
Fig.\ 5: Left:
Apparent velocity $\beta_{app}$ (top)
and position angle (bottom) of new ejected jet components 
plotted versus time of ejection $t_0$. On top, open circles show
the velocity after removal of an
overall slope of $d\beta_{app}/dt= -(0.16 \pm 0.04)$\,yr$^{-1}$.
Superimposed in both figures as a dashed line is a simple precessing 
beam model (cf. Abraham \& Romero, 1999). Whereas the position angle 
follows the model, the apparent velocity shows an additional maximum
near 1988. The data indicate a rotation or precession of the jet base
with a period of 15-16\,yrs. \\
Right: Apparent velocity $\beta_{app}$ of the jet components
plotted versus time of ejection $t_0$ (filled diamonds). Again,
open diamonds show $\beta_{app}$ after removal of the afore mentioned slope.
Superimposed to the velocity is the optical V-Band light-curve
(from T\"urler et al. 1999). The flux density at V-Band is in arbitrary flux units.
In 1982 -- 1997 a correlation between optical flux and ejection velocity
is indicated. 
}
\end{figure}

For many AGN, a correlation between flux density variability and ejection of VLBI components 
is suggested. In 3C\,273 we identified 13 jet components (C6 -- C18) and traced
their motion back to their ejection from the VLBI core. The typical measurement uncertainty
for the ejection times $t_0$ ranges between 0.2 -- 0.5\,yr. In Figure 6 (left) we plot $t_0$
and the flux density variability at 22 -- 230\,GHz. We also add the Gamma-ray
detections of 3C\,273 from EGRET. 
In Figure 6 (right) we plot the onset times of the mm-flares derived
from these light curves (T\"urler et al. 1999) together with the VLBI ejection time ($t_0$) and the 
Gamma-ray fluxes. For each onset of a mm-flare, we find that a new jet component was ejected.

Although the time sampling of the Gamma-ray data is quite coarse,
a relation between component ejection and high Gamma-ray flux 
appears very likely (note that each Gamma-detection 
already means higher than usual $\gamma$-brightness). From a more detailed analysis
(Krichbaum et al. 2002) we obtain for the time lag between component ejection and onset of
a mm-flare: $t_0 - t_0^{\rm mm} = 0.1 \pm 0.2$\,yr. If we assume that the observed peaks in the Gamma-ray
light-curve are located near the times $t_0^\gamma$ of flux density maxima, we 
obtain  $t_0^\gamma -  t_0^{\rm mm} = 0.3 \pm 0.3$\,yr. Although the Gamma-ray variability
may be faster, this result is fully consistent with the more general finding of 
enhanced Gamma-ray fluxes mainly during the rising phase of millimeter flares.
We therefore suggest the following tentative sequence of events:
$t_0^{\rm mm} \leq  t_0 \leq t_0^{\gamma}$ -- the onset of a millimeter
flare is followed by the ejection of a new VLBI component and,
either simultaneously or slightly time-delayed, an increase of the Gamma-ray flux.
If we focus only on those VLBI components, which were ejected close to the main maxima of the 
Gamma-ray light-curve in Figure 6, we obtain time lags of $t_0^{\gamma} - t_0$
of $\leq 0.5$\,yr for C12, $\leq 0.9$\,yr for C13, $\leq 0.2$\,yr for
C16 and $\leq 0.1$\,yr for C18. In all cases, the Gamma-rays seem to
peak a little later than the time of component ejection. With $\beta_{app} \simeq 4$
near the core, the Gamma-rays would then escape at a radius $r_\gamma \leq 0.1$\,mas.
This corresponds to $r_\gamma \leq 2000$ Schwarzschild radii (for a $10^9$ M\solar black hole) 
or $\leq 6 \cdot 10^{17}$\,cm, consistent with theoretical expectations, in which
Gamma-rays escape the horizon of photon-photon pair production at separations  
of a few hundred to a few thousand Schwarzschild radii.
\begin{figure}[t]
\begin{minipage}[t]{6.5cm}{
\psfig{figure=kri_fig6a.epsi,width=5.05cm,angle=-90} 
}
\end{minipage}
~~
\begin{minipage}[t]{6.5cm}{
\psfig{figure=kri_fig6b.epsi,width=4.95cm,angle=-90}
}
\end{minipage}
{\footnotesize ~~\\ Fig.\ 6:
Left: Flux density variations at 230 GHz (filled diamonds), 86 GHz (open circles)
and 22 GHz (filled squares). 
Upward oriented triangles denote Gamma-ray fluxes from EGRET,
downward oriented triangles are upper limits. The extrapolated
ejection times of the VLBI components and their uncertainties
are indicated by filled circles with horizontal bars along the time axis.
Right: Broad-band flux density activity and component ejection.
VLBI component ejection (open circles), Gamma-ray fluxes (triangles, 
downward oriented for upper limits) and onset-times for the millimeter flares 
(open squares, from T\"urler et al. 1999) are shown versus time. 
Labels denote the component identification from VLBI.
}
\vspace{-0.5cm}
\end{figure}
\section{The Contribution of the SRT}
To date, VLBI at 3\,mm and shorter wavelengths is still limited in sensitivity.
Presently, a few  dozens of compact objects with typical flux densities $S_{\rm 86\,GHz} \geq 0.5-1$\,Jy
can be reliably imaged (e.g. Lobanov et al. 2000).
This limitation can be overcome by (i) adding more collecting area, (ii)
by enlarging the observing bandwidth and data recording rate (presently 
$\leq 256$\,Mbits/sec and $\leq 128$\,MHz), and (iii) by correcting 
phase fluctuations introduced by the atmosphere (water vapor radiometry, dual frequency VLBI). 
Another limitation is the non-uniformity of the uv-coverage, in particular the lack of short 
interferometer baselines in present 3\,mm-VLBI. Since many AGN are partially resolved and
show relatively long VLBI jets even at millimeter wavelengths (at 86\,GHz
AGN jets can extend over more than 200 beam sizes, cf. Krichbaum et al. 1999), 
the addition of large and sensitive millimeter telescopes at moderate separations from the existing
telescopes is of considerable importance for reliable images.
In this context and not only because of its large collecting area, but also  
due to its southern location within Europe, could the Sardinia Radio Telescope 
(SRT) make a significant contribution to the scientific research of compact galactic and
extragalactic radio sources at millimeter wavelengths.
\begin{figure}[t]
\begin{minipage}[t]{6.5cm}{
\psfig{figure=kri_fig7a.epsi,width=5.5cm,angle=-90}
}
\end{minipage}
~~
\begin{minipage}[t]{6.5cm}{
\psfig{figure=kri_fig7b.epsi,width=5.5cm,angle=-90}
}
\end{minipage}
{\footnotesize ~~\\ Fig.\ 7:
A simulation of a future uv-coverage for 3C\,273 with the largest millimiter 
antennas in Europe.  On the left the uv-coverage is shown for Bonn, Pico Veleta, 
Plateau de Bure (PDB) and Yebes.  On the right the Sardinia Radio Telescope (SRT) is added.}
\end{figure}

At 86\,GHz, the detection threshold on a single VLBI baseline (7\,$\sigma$, 256\,Mbits/sec, 
60\,sec integration time) is at present 50--100\,mJy on the most sensitive baselines (eg.
Pico Veleta to Effelsberg) and 300--400\,mJy on VLBA baselines. With the 
telescopes summarized in the top section of Table 1, 
the baseline sensitivities within Europe are presently a factor of 2--3 higher
than in America. This will change when the Green Bank Telescope (GBT) and the
Mexican Large Millimeter Telescope (LMT) start observing at 3\,mm. In the near
future, also the phased IRAM interferometer on Plateau de Bure (six 15\,m antennas) and
the 40\,m antenna in Yebes could participate in mm-VLBI. This should 
lower the detection threshold for mm-VLBI in Europe to 30-40\,mJy (at 86\,GHz). The combination
of Plateau de Bure, Yebes and the SRT with the other existing millimeter
antennas (Table 1), will not only improve this sensitivity by another factor of $1.5-2$.
Equally important, this will also ameliorate the uv-coverage on short uv-spacings (see Fig.\ 7),
in particular for sources which are near or below the celestial equator, and which will be 
the prime targets for global mm-VLBI with ALMA. With ALMA, the single baseline sensitivity
between ALMA, the SRT and other large millimeter telescopes will be as low  $5-10$\,mJy.
Under the assumption of future enhanced bandwidth and Giga-bit recording and with full atmospheric phase
correction applied, it could even reach the sub-milli Jansky level. So future mm-VLBI could 
reach the same sensitivity and could observe the same number of compact sources 
as cm-VLBI does nowadays.

\acknowledgments
Many of the results shown here originate from milli\-meter-VLBI 
observations with the following observatories: 
Effelsberg (MPIfR), Onsala and Sest (OSO), Pico Veleta (IRAM), Haystack (MIT-NEROC),
Hat Creek (BIMA), Owens Valley (OVRO), and the VLBA (NRAO).
We like to thank all the people who contributed to these observations
for their efforts and for their continuing enthusiasm to push VLBI to higher frequencies.
TPK also likes to thank the organizers of the SRT-conference for their support and hospitality.

%

\end{document}